\documentclass[10pt,tightenlines,showpacs,letterpaper,aps,prd,longbibliography,twocolumn,nofootinbib,superscriptaddress]{revtex4-1}

\usepackage{graphicx}
\usepackage{amssymb}
\usepackage{epstopdf}
\usepackage{amsmath,amsthm}	
\usepackage[normalem]{ulem} 
\usepackage{comment}
\usepackage{wasysym}
\usepackage{subfigure}
\usepackage{hyperref}
\usepackage{placeins}

\newcommand{\COMMENT}[1]{}
\newtheorem{theorem}{Theorem}[section]
\newtheorem{corollary}{Corollary}[theorem]

\newtheorem{proposition}[theorem]{Proposition}

\newcommand{\beq}{\begin{equation}}
\newcommand{\eeq}{\end{equation}}
\newcommand{\bea}{\begin{align}}
\newcommand{\eea}{\end{align}}
\newcommand{\bq}{\begin{quote}}
\newcommand{\eq}{\end{quote}}

\newcommand{\rob}{\color{black}}

\newcommand{\blk}{\color{black}}

\usepackage{xparse}

\usepackage[x11names]{xcolor}
\ExplSyntaxOn
\newcommand{\colorstring}[3]{%
	\str_set:Nn \l_tmpa_str {#3}
	\int_step_inline:nnnn {1} {1} {\str_count:N \l_tmpa_str } {%
		\int_if_odd:nTF{##1}{
			\textcolor{#1}{\str_item:Nn \l_tmpa_str {##1}}
		}{
			\textcolor{#2}{\str_item:Nn \l_tmpa_str {##1}}
		}%
	}%
}
\ExplSyntaxOff

\begin{document}

\title{Aspects of the phenomenology of interference that are genuinely nonclassical}
\author{Lorenzo Catani}\email{lorenzo.catani@tu-berlin.de}\affiliation{Electrical Engineering and Computer Science Department, Technische Universit\"{a}t Berlin, 10587 Berlin, Germany}

\author{Matthew Leifer}
\affiliation{Institute for Quantum Studies and Schmid College of Science and Technology, Chapman University, One University Drive, Orange, CA, 92866, USA}

\author{Giovanni Scala}
\affiliation{International Centre for Theory of Quantum Technologies, University of Gdansk, 80-308 Gdansk, Poland}

\author{David Schmid}
\affiliation{International Centre for Theory of Quantum Technologies, University of Gdansk, 80-308 Gdansk, Poland}

\author{Robert W. Spekkens}
\affiliation{Perimeter Institute for Theoretical Physics, 31 Caroline Street North, Waterloo, Ontario Canada N2L 2Y5}

\begin{abstract}
Interference phenomena are often claimed to resist classical explanation. However, such claims are undermined by the fact that the specific aspects of the phenomenology upon which they are based can in fact be reproduced in a
noncontextual ontological model [\href{https://quantum-journal.org/papers/q-2023-09-25-1119/}{Catani {\em et al.} Quantum 7, 1119 (2023)}]. 
This raises the question of what {\em other} aspects of the phenomenology of interference {\em do} in fact resist classical explanation. We answer this question by demonstrating that the  most basic 
 quantum wave-particle duality relation, which expresses the precise trade-off between path distinguishability and fringe visibility, cannot be reproduced in any noncontextual model. We do this by showing that it is a specific type of uncertainty relation, and then leveraging a recent result establishing that noncontextuality restricts the functional form of this uncertainty relation [\href{https://doi.org/10.1103/PhysRevLett.129.240401}{Catani {\em et al.} Phys. Rev. Lett. 129, 240401}].  Finally, we discuss what sorts of interferometric experiment can demonstrate contextuality via the wave-particle duality relation. 
\end{abstract}

\maketitle

\section{Introduction}

\begin{figure}
	\centering
	\includegraphics[width=0.45\textwidth]{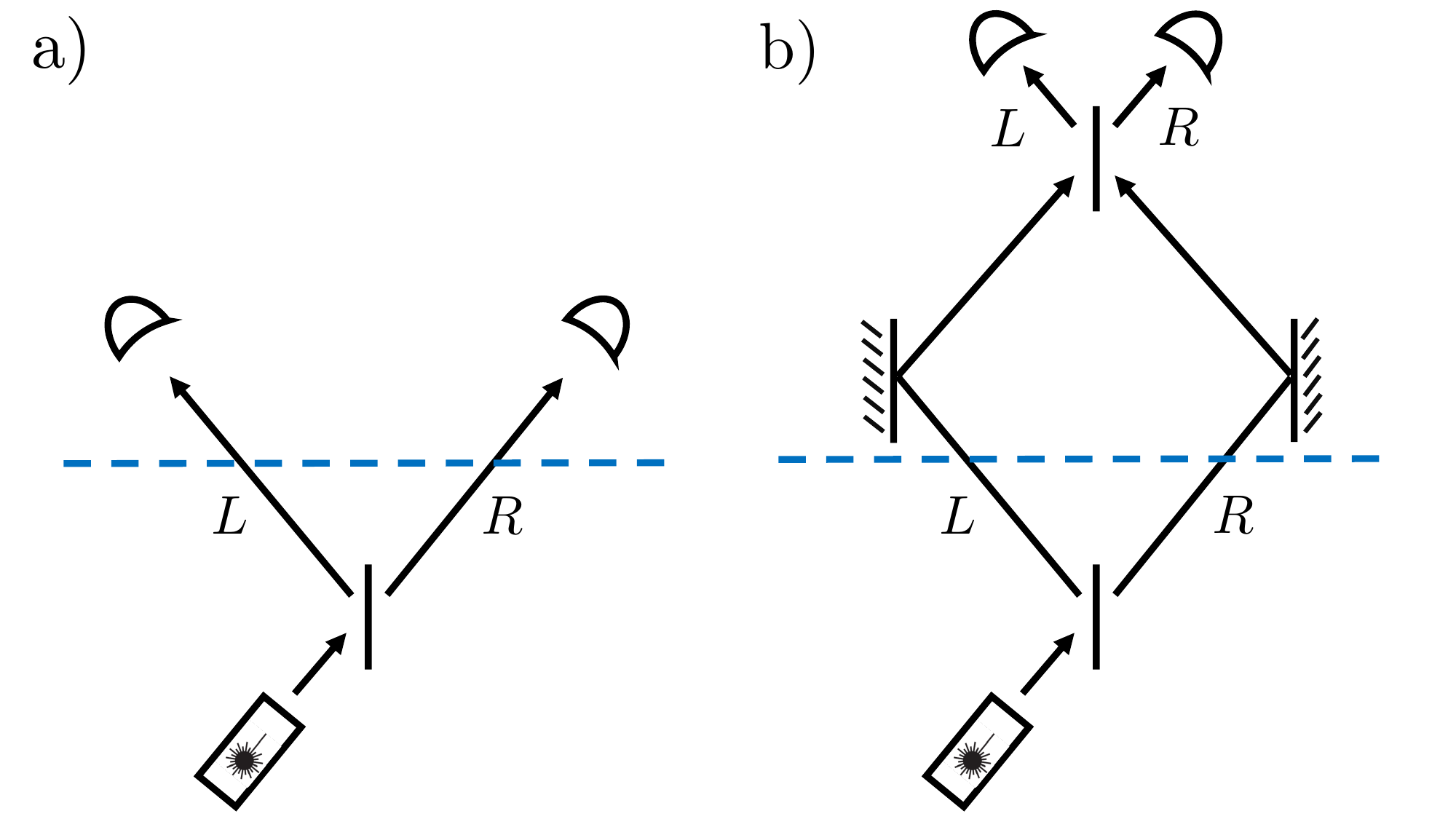}
	\caption{Einstein's thought experiment concerning wave-particle duality.  a) A single photon source at the left input of a 50-50 beamsplitter leads to detection in one and only one arm.  b) If the detectors on the arms are not present and the two beams are recombined at a second 50-50 beamsplitter, interference is observed.  
	}
	\label{MZI}
\end{figure}

The canonical thought experiment concerning wave-particle duality,  as with so many thought experiments in the foundations of quantum theory, originates with Einstein.  Bohr attributes it to him in Ref.~\cite{Bohr1949}, and describes it thus:

\begin{quote}
	If a semi-reflecting mirror is placed in the way of a photon, having two possibilities for its direction of propagation, the photon may either be recorded on one, and only one, of two photographic plates [...]
	in the two directions in question, or else we may, by replacing the plates by mirrors, observe effects exhibiting an interference between the two reflected wave-trains. 
\end{quote}
This pair of possibilities is depicted in Figs.~\ref{MZI}~(a) and ~\ref{MZI}~(b), respectively. The semi-reflecting mirror can be presumed to be a 50-50 beamsplitter, and in the case where the mirrors are in place, the interference is observed by recombining the beams at a second 50-50 beamsplitter.  This set-up, of course, is the famous Mach-Zehnder interferometer~\cite{Zehnder1891,Mach1892}.

In modern versions of the thought experiment, it is usually further noted that (i) if the photographic plates in Fig.~\ref{MZI}~(a) are replaced by nondestructive detectors 
and the beams are again recombined at the second 50-50 beamsplitter (so that one has a Mach-Zehnder interferometer wherein a nondestructive which-way measurement is implemented), then one {\em still} does not see any interference, and (ii) as it is assumed that the source generates only a single photon, \blk it suffices to have a detector in just one of the arms to implement a which-way measurement and destroy the interference.   Consequently, the modern version of the thought experiment differs from that depicted in Fig.~\ref{MZI} in that it typically considers two configurations of a Mach-Zehnder interferometer:  the first, which replaces the set-up depicted in Fig.~\ref{MZI}~(a), is an interferometer wherein there is a (nondestructive) detector on one of the arms between the pair of beamsplitters; 
the second is a variant of Fig.~\ref{MZI}~(b) in which there is a phase shifter on one arm, and where it is sufficient to consider two choices of the local phase shift that allow one to toggle the relative phase shift between 0 and $\pi$. 
 In the second configuration, one sees perfect interference, with the dark port (the one corresponding to destructive interference) toggling between left and right as one toggles the choice of the local phase shift. 
In the first configuration, the presence of the detector destroys this interference. 
	
	Some version of this thought experiment is familiar to all students of quantum theory, since certain aspects of its phenomenology have been heralded as capturing the essence of quantum theory, with Feynman being one of the most prominent advocates of this view~\cite{Feynman1961}. (Note that Feynman considers the case of the double-slit experiment, but the aspects of the phenomenology that he makes use of are features that the double-slit and the Mach-Zehnder interferometer have in common.)  
The relevant aspects of the phenomenology include those in the previous paragraph (and some extensions thereof); see Ref.~\cite{ToyFieldTheory} for a complete discussion. 
	In that article, the collection of these aspects is called the {\em TRAP phenomenology}, where TRAP stands for ``traditionally regarded as problematic''.  
	
	The claim that the TRAP phenomenology captures the essence of quantum theory is {\em unlike} similar claims that focus on Bell inequality violations~\cite{Bell_1964,bell1995theory,Clauser1969,RevModPhys.86.419}
	or violations of inequalities derived from generalized noncontextuality~\cite{Spekkens2005,POM,Ravi1,Ravi2,Schmid2018all,Chaturvedi2021}    because it is not supported by a no-go theorem that establishes rigorously the inconsistency between the TRAP phenomenology and certain stipulated principles  that formally define a notion of classical explainability (such as the principles of locality or noncontextuality within the framework of ontological models~\cite{Harrigan2010}). This leaves open the possibility that a classical model of this interference phenomenology could be constructed.  Indeed, such a model {\em does} exist, as was shown in Ref.~\cite{ToyFieldTheory}.  
	This model is classical in the sense that its kinematical state space is represented by a set and its dynamics are represented by 
	functions on this set~\cite{Schmid2021unscrambling}, while also respecting the principles of locality and generalized noncontextuality. 
	
	We here take a  phenomenon to be classically explainable  if and only if it can be reproduced in a generalized-noncontextual ontological model~\cite{Spekkens2005}. We define this notion formally further on, but pause here to make a few comments about its credentials. Generalized noncontextuality can be motivated by a methodological version of Leibniz's principle of the ontological identity of empirical indiscernibles~\cite{Leibniz, Schmid2021unscrambling}, or by the principle of operational no fine-tuning~\cite{Catani2023}. 
	Realizability by a generalized-noncontextual ontological model 
implies
 the existence of a locally causal model~\cite{Bell_1964,bell1995theory}, the existence of a Kochen-Specker-noncontextual model~\cite{KochenSpecker1967} for sharp measurements~\cite{determinism,Ravi1}, the existence of a simplex-embedding for a generalized probabilistic theory~\cite{SchmidGPT,selby2021accessible}, and the existence of a nonnegative quasiprobability representation~\cite{schmid2020structure,Spekkens2008}, and is implied by the existence of a macrorealist explanation~\cite{schmid2022macrorealism}. 
Under appropriate conditions, various other indicators of nonclassicality (including anomalous weak values~\cite{Pusey2014}, violations of the Leggett-Garg inequality~\cite{forthcomingLGI}, and universal quantum computation~\cite{Howard,schmid2022uniqueness}) imply the failure of generalized noncontextuality. The failure of generalized noncontextuality is also a resource for various forms of information processing~\cite{POM,RAC,RAC2,Saha_2019,SchmidSpekkens2018,lostaglio2020contextual,SahaAnubhav2019,Tavakoli2017,Yadavalli2020,Roch2021,Flatt2021}.	
	
	Having a precise and well-motivated boundary between classicality and nonclassicality is important because it informs our attempts to solve various problems:
	(i) the project of extending
	quantum theory into new domains, such as gravitational physics, and of proposing quantum versions of
	various classical theoretical frameworks, 
	such as the frameworks of causal inference~\cite{Pearl}
	or algorithmic information theory\cite{Li2019,Chaitin2004};
	(ii) the project of uncovering novel quantum-over-classical advantages for various tasks
	in computation, communication, and cryptography \cite{nielsen_chuang_2010}; (iii) the project of determining the correct
	interpretation of the quantum formalism.

	Given that the TRAP phenomenology of interference can be reproduced in a generalized-noncontextual ontological model, and hence is classically explainable, 
	it is natural to ask: what aspects of interference phenomena, if any, {\em are} genuinely nonclassical? That is: what aspects of interference phenomena {\em cannot} be reproduced by any generalized-noncontextual ontological model? \blk
	
	An easy but relatively uninformative way to answer this question is to take any known proof of the failure of 
	generalized noncontextuality and recast it into an interferometric scenario.  For instance, a proof involving one or two abstract qubits can be instantiated interferometrically by associating each qubit with the two-dimensional Hilbert space corresponding to the which-path degree of freedom of the system, termed a {\em dual-rail qubit}.
	This can always be done, but one learns very little in the process, since it is merely a restatement of known facts about the Hilbert space structure of quantum theory in the language of interferometers. 
	The more difficult but more informative way to answer the question is to find some 
operational phenomenology that was previously thought to be significant {\em to the study of quantum interference} 
 and show that this 
 phenomenology witnesses nonclassicality,  in the sense of being inconsistent with a generalized-noncontextual or local ontological model. 
	
	We show herein that the functional form of the most basic
	wave-particle duality relation in quantum theory~\cite{greenberger1988simultaneous} is such an aspect.
	
\begin{figure}[htbp] 
		\centering
		\includegraphics[width=.49\textwidth]{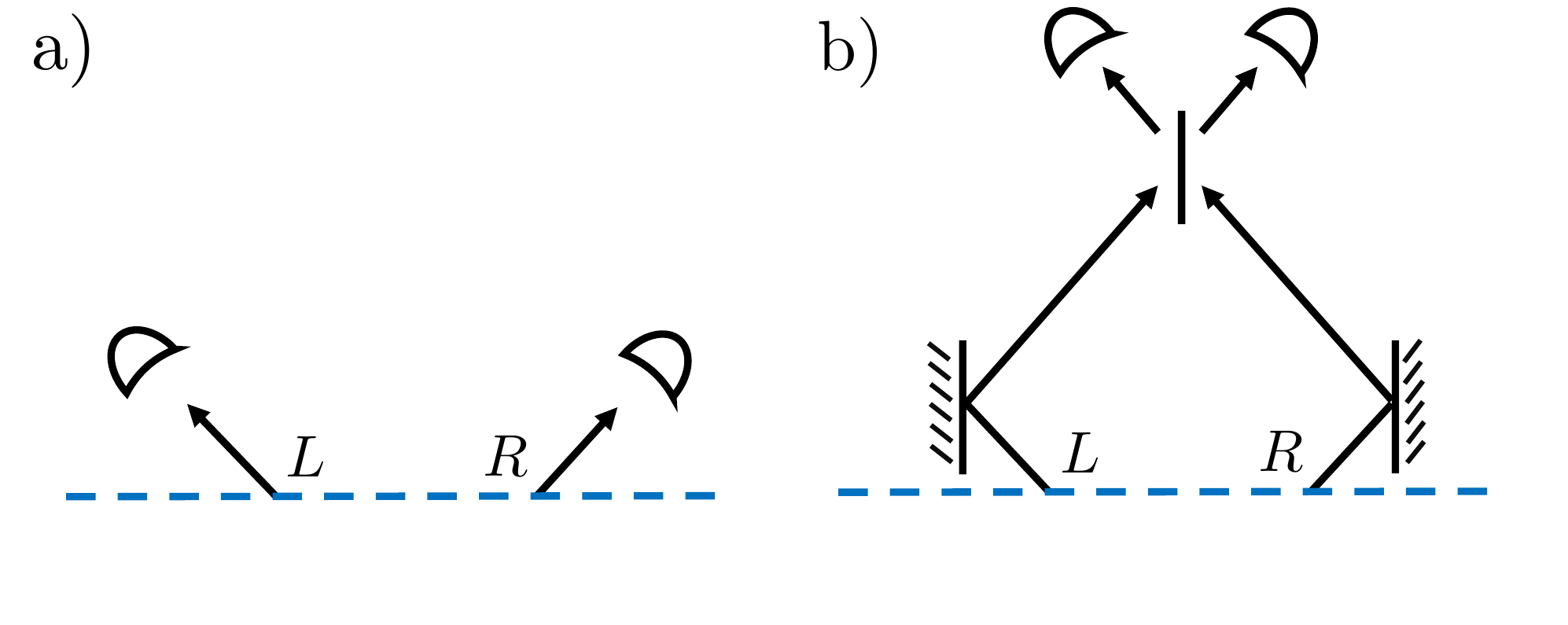} 
		\centering
		\includegraphics[width=.29\textwidth]{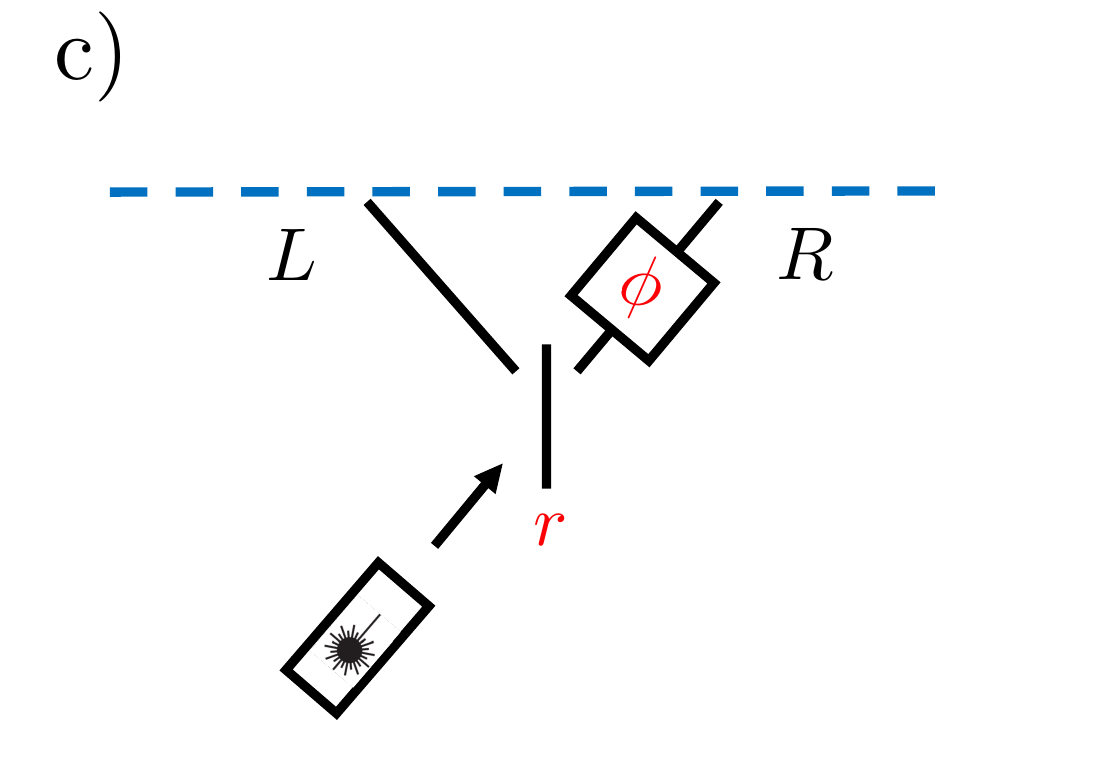} 
		\caption{ Measurement and preparation stages of the interferometric experiment. a) Which-way and b) which-phase measurements. c) A preparation procedure with beamsplitter reflectivity $r$ and phase shift $\phi$ allowing for the implementation of any quantum state of the dual-rail qubit. 
		}
		\label{MeasPrep}
	\end{figure}

A {\em wave-particle duality relation}	is a tradeoff relation between two quantities that are meant to represent wave-like and particle-like behaviours respectively.  For the relation of interest to us here~\cite{wootters1979complementarity,greenberger1988simultaneous}, these quantities are the {\em fringe visibility} $\mathcal{V}$ and the {\em path distinguishability} $\mathcal{P}$. The path distinguishability is defined in terms of the statistics of a measurement of the type depicted in Fig.~\ref{MeasPrep}~(a), which we term a {\em which-way measurement}, while the fringe visibility is defined in terms of the statistics of a measurement of the type depicted in  Fig.~\ref{MeasPrep}~(b), which we term a {\em which-phase measurement}.  (Their precise definitions are given further on.) If we conceptualize the two experiments depicted in Fig.~\ref{MZI} as involving a preparation stage and a measurement stage for the dual-rail qubit (the division between the stages being indicated by the blue dashed line), \rob then we see that the two possibilities for the measurement stage are what we have picked out in Figs.~\ref{MeasPrep}~(a) and (b). \blk

 The tradeoff relation we will consider between the fringe visibility $\mathcal{V}$ and path distinguishability $\mathcal{P}$ holds for {\em all} states of the dual-rail qubit.
One of the ways of preparing an arbitrary state of the dual-rail qubit is depicted in Fig.~\ref{MeasPrep}~(c), where the beamsplitter is allowed to have any reflectivity $r$ in the range $[0,1]$ and there is a phase shifter on one of the arms that can implement any phase shift $\phi$ in the range $[0,2\pi)$.  Thus, for each preparation of Fig.~\ref{MeasPrep}~(c), we can imagine following it up with the measurement in Fig.~\ref{MeasPrep}~(a) or the measurement in Fig.~\ref{MeasPrep}~(b), and it is the statistics of these two counterfactual possibilities for the measurement that are constrained by the wave-particle duality relation of interest to us. \blk

\begin{figure}[htbp] 
		\centering
		\includegraphics[width=.5\textwidth]{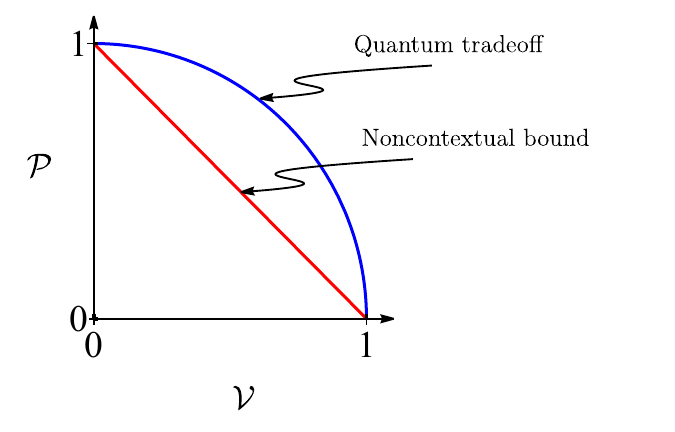} 
		\caption{Plots of fringe visibility $\mathcal{V}$ versus path distinguishability $\mathcal{P}$.  The blue curve is the quantum tradeoff relation  (Eq.~\eqref{QTradeoff}). The red curve is the bound on the tradeoff relation that is implied by the assumption that there exists a noncontextual ontological model of the experimental set-up (Eq.~\eqref{NCBoundTradeoff}). }
		\label{QTradeoffRelationVsNCBound}
	\end{figure}

We can now summarize our main result. 
 Under a particular symmetry condition (which we discuss below and which always holds in quantum theory), noncontextuality implies the constraint 
	\begin{equation}\label{NCBoundTradeoff}
		\mathcal{V}+\mathcal{P} \le 1.
	\end{equation}
	This constitutes a noncontextuality inequality that is noise-robust in the sense of Refs.~\cite{Mazurek2016,Ravi1,Pusey2018robust}.
	Meanwhile, the quantum tradeoff between visibility $\mathcal{V}$ and distinguishability $\mathcal{P}$ is described by the relation~\cite{wootters1979complementarity,greenberger1988simultaneous} 
	\begin{equation}\label{QTradeoff}
		\mathcal{V}^2+\mathcal{P}^2\le 1.
	\end{equation}
In Fig.~\ref{QTradeoffRelationVsNCBound}, this tradeoff is plotted alongside the noncontextual bound of Eq.~\eqref{NCBoundTradeoff}. It is clear from this plot that the noncontextual bound can be violated in quantum theory, demonstrating the impossibility of explaining this aspect of quantum phenomenology in terms of a noncontextual ontological model.

	It is interesting to note that Greenberger and Yasin~\cite{greenberger1988simultaneous} 
	explicitly expressed surprise at how large the fringe visibility can be for a given degree of path distinguishability 
	in the quantum tradeoff relation of Eq.~\eqref{QTradeoff}.  They did not, however, articulate their reasons for expecting otherwise. While it is true that the {\em simplest} form of such a tradeoff would be linear (so that an increase in path distinguishability implies a {\em proportional} decrease in fringe visibility), this is not a compelling reason to expect linearity.  
	Our result, however, provides such a reason: if it were possible to explain the experimental set-up classically, in the sense of reproducing its predictions within a generalized-noncontextual ontological model, 
then 	the tradeoff would be linear.

\section{The wave-particle duality relation}

	The fringe visibility $\mathcal{V}$ in a generic interference pattern is defined as the normalized difference between the maximum and minimum intensities of the pattern:
	\begin{equation}
		\mathcal{V}= \frac{I_{\mathrm{max}}-I_{\mathrm{min}}}{I_{\mathrm{max}}+I_{\mathrm{min}}}.
	\end{equation}
	For the type of interference that is of interest in quantum theory, the pattern is built up from detections of individual particles that exhibit statistical variation in their location (rather than, for instance, describing a continuous physical quantity, such as the height of a water wave).  The intensity at a given location tracks the number of detections in 
	that location  and is consequently proportional to the probability of finding an individual particle there, so that we can replace intensities with probabilities in the expression for $\mathcal{V}$, 
	\begin{equation}\label{fringevis}
		\mathcal{V}= \frac{\mathbb{P}_{\mathrm{max}}-\mathbb{P}_{\mathrm{min}}}{\mathbb{P}_{\mathrm{max}}+\mathbb{P}_{\mathrm{min}}}.
	\end{equation}
	The expression further simplifies when we specialize to the case of the measurement depicted in Fig.~\ref{MeasPrep}~(b). 
	In such an interference pattern, there are only two spatial locations of interest, namely, the two output ports of the second beamsplitter.  Consequently, the two probabilities of detection sum to unity, trivializing the denominator. 

	Meanwhile, the path distinguishability $\mathcal{P}$ is defined as the absolute value of the difference in the probability of finding the particle on the left and the probability of finding it on the right \rob for \blk a which-way measurement, such as the one depicted in Fig.~\ref{MeasPrep}~(a), i.e.,
	\begin{align}\label{pathdist}
		\mathcal{P}&= | \mathbb{P}(L) - \mathbb{P}(R) |.
	\end{align}

	Quantum theory predicts that there is a tradeoff between the fringe visibility and the path distinguishability which holds for all states, and hence for all preparations of the type depicted in Fig.~\ref{MeasPrep}~(c), namely, $\mathcal{V}^2+\mathcal{P}^2 \le 1$ (Eq.~\eqref{QTradeoff}).
	This relation was first derived by Wootters and Zurek~\cite{wootters1979complementarity}, who were considering the double-slit experiment, and it was later rederived in the context of the Mach-Zehnder interferometer by Greenberger and Yasin~\cite{greenberger1988simultaneous}.~\footnote{Note that this relation is distinct from the one described by Englert~\cite{Englert1996} (Eq.~(10) therein) which concerns the tradeoff between the fringe visibility and path distinguishability when one performs unsharp versions of the which-phase and which-way measurements {\em simultaneously}, rather than sharp versions of these as counterfactual alternatives. 
	} 
	
The extreme cases in the tradeoff relation of Eq.~\eqref{QTradeoff} are $(\mathcal{V}, \mathcal{P})=(1,0)$ and $(\mathcal{V}, \mathcal{P})=(0,1)$.  The point $(\mathcal{V}, \mathcal{P})=(1,0)$ can be achieved by the preparation procedure where the beamsplitter has reflectivity 1/2 (a 50-50 beamsplitter) and the relative phase-shift between the two paths is 0 or $\pi$, because in this case the which-phase measurement is perfectly predictable while the which-way measurement is completely unpredictable. The point $(\mathcal{V}, \mathcal{P})=(0,1)$ can be achieved by the preparation procedure where the beamsplitter has reflectivity 0 (i.e., no beamsplitter) or reflectivity 1 (a perfect mirror), because in this case it is the which-way measurement that is perfectly predictable and the which-phase measurement that is completely unpredictable. If one wishes to make either of these quantities nonzero, the other must become strictly less than unity.  The particular manner in which this occurs is what the wave-particle duality relation stipulates.	

\section{Why the wave-particle duality relation is an uncertainty relation}\label{WPDRasUR}

As noted above, it is sufficient for our purposes to conceptualize the system that passes between the preparation stage and the measurement stage as a {\em dual-rail qubit}, that is, as a system described by a two-dimensional Hilbert space where the basis $\{ |L\rangle, |R\rangle\}$ denotes the two eigenstates of a which-way measurement, corresponding to the two ``rails'' or spatial modes of the interferometer on which the particle might be found; here, $L$ denotes the left mode and $R$ denotes the right mode.\footnote{This is a first-quantized description of the experimental set-up.  In Ref.~\cite{ToyFieldTheory}, it is argued that in order to properly analyze questions about locality, the first-quantized description should be understood to concern coarse-grainings of the excitational degrees of freedom of field modes, rather than the motional degrees of freedom of particles.  Because we do not concern ourselves with issues of locality here, we do not emphasize the distinction.}

The which-way measurement depicted in Fig.~\ref{MeasPrep}~(a) corresponds to measuring  the dual-rail qubit in the $\{ |L\rangle, |R\rangle\}$ basis. The Hermitian operator associated with such a measurement when one associates the $|L\rangle$ ($|R\rangle$) outcome with the eigenvalue $+1$ ($-1$) is given by the 
operator $\hat{Z} := |L\rangle \langle L| -|R\rangle \langle R|$,  that is, the Pauli-$Z$ operator relative to the $\{ |L\rangle, |R\rangle\}$ basis. 
We term this the  {\em which-way observable}. 
Letting $\mathbb{P}_Z$ denote the probability distribution over the outcome of this measurement (a random variable which we denote by $Z$), the path distinguishability $\mathcal{P}$ defined in Eq.~\eqref{pathdist} can be expressed as
\begin{align}\label{www}
	\mathcal{P}&= | \mathbb{P}_{Z}(+1) - \mathbb{P}_{Z}(-1)|\nonumber\\
	&=|\langle Z \rangle|.
\end{align}
Here, $\langle Z\rangle = {\rm tr}(\hat{Z} \rho)$, where $\rho$ denotes the quantum state of the dual-rail qubit associated to the preparation stage.\blk 

For a measurement with outcome $M$ in the set $\{+1,-1\}$, the absolute value of its expectation value, $|\langle M\rangle|$, is a measure of the predictability of the measurement. Its maximum value $|\langle M\rangle|=1$  occurs for a distribution that is weighted entirely on the outcome +1 or entirely on the outcome $-1$, and its minimum value $|\langle M\rangle|=0$ occurs for a distribution that is uniform over the two outcomes. 

Eq.~\eqref{www} therefore shows that  the path distinguishability $\mathcal{P}$ is simply a measure of the predictability of the which-way measurement, \rob and it is consequently appropriate to describe \blk
$\mathcal{P}$ as the 
{\em which-way predictability}.

We now consider the fringe visibility. 
Note that in its definition, Eq.~\eqref{fringevis}, the maximum and minimum probabilities are obtained by ranging over all the values of the phase shift on one arm.  These extrema are achieved for a pair of phases, $\phi_{\rm max}$  and $\phi_{\rm  max}+\pi$. For convenience, we henceforth consider an interferometer wherein the path-lengths are chosen such that it is possible to take $\phi_{\rm  max}=0$ and that for this trivial phase shift, it is the detector at the left output port of the second beamsplitter that fires. 

For such an interferometer, the beamsplitter transformation in the measurement depicted in Fig.~\ref{MeasPrep}~(b) is represented by the Hadamard unitary (relative to the basis $\{ |L\rangle, |R\rangle\}$) and therefore the effective measurement on the dual-rail qubit is \rob of \blk the basis 
$\{ |+\rangle, |- \rangle \}$, where the two states
$| \pm \rangle := \tfrac{1}{\sqrt{2}}(|L\rangle \pm |R\rangle)$ 
correspond to the relative phase between the left and right modes being $0$ or $\pi$ respectively.  
It is for this reason that it is apt to call the measurement depicted in Fig.~\ref{MeasPrep}~(b) a {\em which-phase} measurement.   \rob We adopt the convention that the outcome $+1$ ($-1$) is associated to the firing of the detector at the left (right) output port of the second beamsplitter, and hence to the case where the relative phase is 0 ($\pi$).  \blk
 In this case, the Hermitian operator associated to this measurement, the {\em which-phase observable}, is $\hat{X} = | + \rangle \langle +| -| - \rangle \langle - |$, \rob that is, the Pauli-$X$ operator relative to the $\{ |L\rangle, |R\rangle\}$ basis. \blk
  Letting $\mathbb{P}_X$ denote the probability distribution over the outcome of this measurement (a random variable which we denote by $X$), the fringe visibility $\mathcal{V}$ defined in Eq.~\eqref{fringevis} can be expressed as
\begin{align}
	\mathcal{V}&=  \max \{\mathbb{P}_{X}(+1),\mathbb{P}_{X}(-1)\} - \min \{\mathbb{P}_{X}(+1),\mathbb{P}_{X}(-1)\}\nonumber\\
	&= | \mathbb{P}_{X}(+1) - \mathbb{P}_{X}(-1)|\nonumber\\
	&= |\langle {X} \rangle|.\label{Vsimple}
\end{align}
 \blk

Eq.~\eqref{Vsimple} shows that the fringe visibility $\mathcal{V}$ is simply the absolute value of the expectation value of the $\pm 1$-valued variable $X$ that describes the outcome of the which-phase measurement, and so is a measure of the {\em predictability} of this measurement. 
 \rob It is consequently appropriate to describe \blk $\mathcal{V}$ as the {\em which-phase predictability}.

Given the identification of $\mathcal{V}$ and $\mathcal{P}$ as predictabilities of the outcomes of the which-phase and which-way measurements respectively, the wave-particle duality relation of Eq.~\eqref{QTradeoff} can be understood as a tradeoff of predictabilities, or equivalently, as a tradeoff of unpredictabilities or uncertainties.  
An {\em uncertainty relation} is a tradeoff of predictabilities that holds for {\em all states}.   Because the tradeoff between which-path and which-way predictability is of this type,
it describes an uncertainty relation. 

In fact, the wave-particle duality relation is simply a special case of an uncertainty relation for an arbitrary qubit. 
One can quantify the predictability of  measurements of Pauli-$Z$ and Pauli-$X$ observables by $|\langle Z\rangle|$ and  $|\langle X\rangle|$ respectively.  \rob It follows from the fact that the quantum state space of a qubit is the Bloch sphere that \blk 
  the tradeoff relation that holds between these predictabilities is
\begin{equation}\label{QuantumZXUR}
	\langle Z\rangle^2 + \langle X\rangle^2 \le 1.
\end{equation} 
\rob This can be understood as an uncertainty relation for a qubit.  (See Appendix D
 of Ref.~\cite{ContextualityViaUR} for an account of the history of conceptualizing Eq.~\eqref{QuantumZXUR} as an uncertainty relation.) \blk
Given Eqs.~\eqref{www} and \eqref{Vsimple}, \rob we see that the quantum wave-particle duality relation of Eq.~\eqref{QTradeoff} is simply an instance of this uncertainty relation, \blk  
specialized to the case of the which-way and which-phase observables of a dual-rail qubit.
\blk

It should be noted that Greenberger and Yasin~\cite{greenberger1988simultaneous} explicitly {\em denied} that their wave-particle duality relation could be interpreted as an uncertainty relation.  Our impression is that this arose from having an overly narrow conception of what constitutes an uncertainty relation.  
Many other authors, however, {\em have} recognized that wave-particle duality relations wherein the which-way and which-phase measurements are counterfactual alternatives
  (the type considered here) are instances of uncertainty relations~\cite{jaeger1995two,bjork1999complementarity,luis2001complementarity,liu2012relation,Coles2014,Coles2017}.

\section{Operational theories, ontological models and generalized noncontextuality}

We now briefly introduce the relevant preliminaries for discussing noncontextuality for 
prepare-measure scenarios. For a given system, an {\em operational theory} stipulates the possible preparations and measurements, as well as the probability $\mathbb{P}(y|M,P)$ of obtaining each outcome $y$ of measurement $M$ when performed on preparation $P$. When characterized in the framework of generalized probabilistic theories (GPTs)~\cite{hardy2001quantum,barrett2007information,chiribella2010probabilistic}, one represents each preparation $P$ and each effect $[y|M]$ by real-valued vectors $\vec{s}_P$ and $\vec{e}_{y|M}$, respectively, where the probabilities are given by $\mathbb{P}(y|M,P)= \vec{s}_P \cdot \vec{e}_{y|M}$. 

An {\em ontological model} of an operational theory provides an explanation of the operational predictions of the latter in terms of processes (stochastic maps) on an underlying classical state space~\cite{Harrigan2010}. An ontological model associates a set $\Lambda$, known as the ontic state space, with each given system. (For our purposes this set can be taken to be finite without loss of generality.) An element $\lambda \in \Lambda$ is known as an ontic state, and encodes all physical properties of the system. The ontological model represents each preparation $P$ in the operational theory as a probability distribution $\mu(\lambda |P)$ over ontic states.  Each effect $[y|M]$ is represented by a conditional probability distribution $\xi(y|M,\lambda)$ describing  the probability of obtaining outcome $y$ given that measurement $M$ was implemented on a system with ontic state $\lambda$. We can associate $\mu(\lambda|P)$ and $\xi(y|M,\lambda)$  with vectors,  denoted (respectively) by $\vec{\mu}_P$ and $\vec{\xi}_{y|M}$.  
An ontological model of an operational theory must reproduce the predictions of that theory as follows, 
\begin{align} \label{opdata}
	\mathbb{P}(y|M,P)
	&=\sum_{\lambda\in\Lambda} \xi(y|M,\lambda)\mu(\lambda |P)= \vec{\xi}_{y|M} \cdot \vec{\mu}_P.
\end{align}
Such a model satisfies
 the principle of generalized noncontextuality\footnote{Henceforth, we shall use the term ``noncontextuality'' as a shorthand for ``generalized noncontextuality''.} if {\em operationally equivalent} laboratory procedures are represented as identical  processes in the ontological model~\cite{Spekkens2005}.  
In the case of preparation procedures (which is all we require in this work), this can be formalized as follows.
Two preparation procedures $P$ and $P'$ are operationally equivalent if they lead to the same statistics for all possible measurements, so that $\forall M: \mathbb{P}(y|M,P)= \mathbb{P}(y|M,P')$. We denote this equivalence relation by $P\simeq P'$. Then, an ontological model is noncontextual if any two such preparations are represented by the same probability distribution over ontic states:
\begin{equation}
	P\simeq P' \implies \vec{\mu}_P= \vec{\mu}_{P'}.
\end{equation}

The representation of operational procedures as real-valued GPT vectors (introduced above) throws away all information about those procedures except for their operational equivalence class. As a consequence, a noncontextual ontological representation of an operational theory is one wherein all procedures associated to the same GPT state vector are represented by the same probability distribution over ontic states. For example, if two different mixtures of GPT states are equal, the corresponding mixtures of their representations as probability distributions must also be equal:
\begin{align} \label{linearnc}
	\sum_i w_i \vec{s}_{i} = \sum_j w'_j \vec{s}^{\;\prime}_{j}\;\;\implies\;\; \sum_i  w_i \vec{\mu}_i = \sum_j w'_j \vec{\mu}^{\;\prime}_{j},
\end{align}
where $\{w_i\}_i$ and $\{w'_{j}\}_{j}$ are probability distributions.

\section{Witnessing contextuality via the wave-particle duality relation}
\label{Sec:ContextualityViaUR}

\subsection{A noncontextuality no-go theorem based on wave-particle duality}

 Our main result is an application of
 a result from Ref.~\cite{ContextualityViaUR} concerning how to witness contextuality via uncertainty relations. We therefore begin by summarizing this prior result.

Consider an arbitrary operational theory and two binary-outcome measurements therein, denoted $M$ and $M'$. We will also use $M$ and $M'$ to denote the outcomes of these measurements, which are assumed to 
take values in the set $\{+1,-1\}$.  

We will consider the tradeoff between the $M$-predictability and the $M'$-predictability for an operational state that satisfies a certain condition, termed  {\em $A_1^2$-orbit realizability} (which is defined relative to $M$ and $M'$).  
It consists of two subconditions. The first is that the operational state has {\em equal-predictability counterparts relative to $M$ and $M'$}: if the operational state is $\vec{s}_1$, then one can find three other operational states, $\vec{s}_2$, $\vec{s}_3$, and $\vec{s}_4$, such that the quadruple of states give equal predictabilities for the $M$ and $M'$ measurements, but vary over all possible signs of the expectation values for the pair:
\begin{align}\label{eq:rectanglesymmetryEV}
&\langle M\rangle_{\vec{s}_1} = \langle M\rangle_{\vec{s}_2} = - \langle M\rangle_{\vec{s}_3} = - \langle M\rangle_{\vec{s}_4},\nonumber\\
& \langle M'\rangle_{\vec{s}_1}= - \langle M'\rangle_{\vec{s}_2} = - \langle M'\rangle_{\vec{s}_3} = \langle M'\rangle_{\vec{s}_4}.
\end{align}
\rob Notice that all four of these states assign the same values to $|\langle M\rangle|$ and $|\langle M'\rangle|$ and therefore any tradeoff between $|\langle M\rangle|$ and $|\langle M'\rangle|$ applies to all of them. \blk
The second subcondition is that this quadruple of states satisfies the {\em operational equivalence relation}
\begin{equation}\label{eq:op_equiv}
\frac{1}{2}\vec{s}_1+ \frac{1}{2}\vec{s}_3 = \frac{1}{2}\vec{s}_2+ \frac{1}{2}\vec{s}_4. 
\end{equation}
(Note that {\em some} operational equivalence relation among the states is always required in order for the principle of noncontextuality to imply constraints on the ontological model.)

In summary, a state satisfies the $A_1^2$-orbit-realizability condition relative to $M$ and $M'$ if it is part  of a quadruple of states satisfying Eqs.~\eqref{eq:rectanglesymmetryEV} and \eqref{eq:op_equiv}.  The condition is referred to as ``$A_1^2$-orbit-realizability'' because any such quadruple of states is an orbit of $A_1^2$, the Coxeter group describing the abstract version of the group of reflections across two planes.

An operational theory is said to have {\em $A_1^2$-symmetry relative to $M$ and $M'$} if {\em all} operational states in the theory satisfy the $A_1^2$-orbit-realizability condition relative to $M$ and $M'$.

\begin{figure}[htbp] 
\centering
\subfigure[]{\includegraphics[width=.246\textwidth]{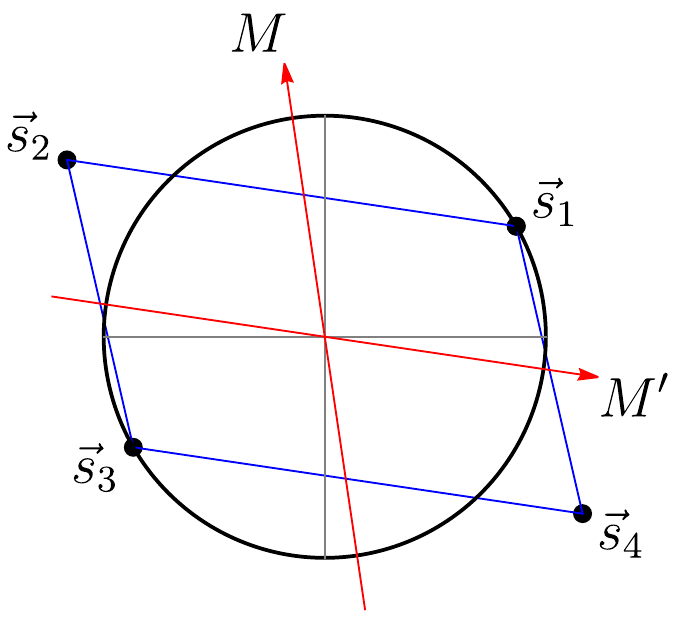}}
\subfigure[]{\includegraphics[width=.231\textwidth]{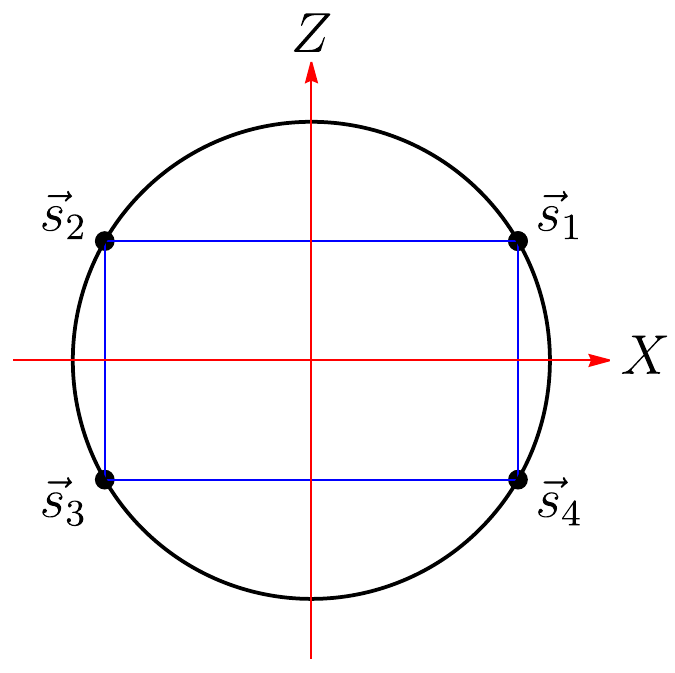}}
\caption{(a) Example of a pair of quantum measurements, $M$ and $M'$, relative to which qubit quantum theory fails to have $A_1^2$-symmetry.  (b) By contrast, for a pair of complementary measurements, such as Pauli $Z$ and Pauli $X$, qubit quantum theory has $A_1^2$-symmetry.}
\label{orbitrealizability}
\end{figure}

For example, consider qubit quantum theory. The only pairs of measurements relative to which {\em all} states satisfy the $A_1^2$-orbit realizability condition are pairs of measurements associated to complementary bases (i.e., orthogonal directions in the Bloch sphere).  This fact can be inferred from the geometry of the Bloch sphere, as we now explain (see Fig.~\ref{orbitrealizability}).  Consider two observables, $M$ and $M'$, whose +1 outcomes are associated to directions that subtend an angle strictly greater than $90^{\circ}$ in the Bloch sphere.  Now consider the Bloch vector $\vec{s}_1$ that describes a pure state that assigns positive expectation values to both $M$ and $M'$.
The state  $\vec{s}_1$ is an example of one that does not satisfy the $A_1^2$-orbit-realizability condition relative to $M$ and $M'$.  This is because, to satisfy the condition, there must be an operational state $\vec{s}_2$ that satisfies $\langle M\rangle_{\vec{s}_2} =\langle M\rangle_{\vec{s}_1}$ and 
$\langle M'\rangle_{\vec{s}_2} =-\langle M'\rangle_{\vec{s}_1}$, but this places $\vec{s}_2$ outside the Bloch sphere, so that it does not describe a valid quantum state. \rob If $M$ and $M'$ describe a pair of observables whose +1 outcomes are associated to directions that subtend an angle strictly {\em less} than $90^{\circ}$ in the Bloch sphere, then any pure state that assigns a positive expectation value to $M$ and a negative expectation value to $M'$ provides an example of one that does not satisfy the $A_1^2$-orbit-realizability condition, by similar logic. In this way, we reach the conclusion that $M$ and $M'$ must correspond to orthogonal directions in the Bloch sphere. For any pair of orthogonal directions (corresponding to a pair of complementary measurements), it is easy to see that every state in the Bloch sphere satisfies the $A_1^2$-orbit-realizability condition, as illustrated in Fig.~\ref{orbitrealizability}(b). 

The which-way and which-phase measurements in quantum theory are an instance of a pair of complementary measurements, and so every state of a dual-rail qubit satisfies $A_1^2$-orbit-realizability relative to the which-way and which-phase measurements.  In other words,  the dual-rail qubit has the $A_1^2$-symmetry property.

The significance of having the $A_1^2$-symmetry property is that it is under this condition that noncontextuality implies a bound on 
 the predictability tradeoff for the pair of measurements in question.  Specifically, the operational equivalence of Eq.~\eqref{eq:op_equiv}, together with the specific instance of noncontextuality described in Eq.~\eqref{linearnc}, implies nontrivial constraints on the ontological representations of the states.  These constraints have consequences for the possible measurement statistics, which are derived in Ref.~\cite{ContextualityViaUR} (see Appendix A), and which we recall here: 
\begin{proposition}\label{ContFromUR}
Consider an operational theory having $A_1^2$-symmetry relative to the pair of measurements $M$ and $M'$.   The operational theory admits of a noncontextual ontological model if and only if the tradeoff of predictabilities for this pair of measurements satisfies the following bound for all states:
\begin{equation}\label{NCBoundZXUR}
|\langle M\rangle| + |\langle M'\rangle| \le 1.
\end{equation}
\end{proposition}

If we take $M$ and $M'$ to be the which-way and which-phase measurements respectively, so that $|\langle M\rangle|= |\langle Z\rangle|=\mathcal{P}$ (where we have used Eq.~\eqref{www}), and $|\langle M' \rangle|= |\langle X\rangle|=\mathcal{V}$ (where we have used Eq.~\eqref{Vsimple}), then we get the following corollary:
\begin{corollary}\label{ContFromWPDuality}
Consider an operational theory having $A_1^2$-symmetry relative to which-way and which-phase measurements.
The operational theory admits of a noncontextual ontological model if and only if the wave-particle duality relation in the operational theory satisfies the  bound 
\begin{equation}\label{NCBoundWPDuality}
\mathcal{V} +\mathcal{P} \le 1.
\end{equation}

\end{corollary}

As we noted above,
a dual-rail qubit has the $A_1^2$-symmetry property relative to the which-way and which-phase measurements.  Consequently, the fact that it can saturate the wave-particle duality relation of Eq.~\eqref{QTradeoff}, namely, $\mathcal{V}^2+ \mathcal{P}^2 \le 1$, and thereby violate the bound of Eq.~\eqref{NCBoundWPDuality}, implies that the functional form of its wave-particle duality relation witnesses contextuality.  This is illustrated in Fig.~\ref{QTradeoffRelationVsNCBound}.

\rob
Imagine now some operational theory that describes an alternative to quantum theory, i.e., a {\em foil} to quantum theory~\cite{Spekkens2016}.  Suppose that it \blk
 allows for a which-way measurement and a which-phase measurement, and that it has $A_1^2$-symmetry relative to these measurements. \rob Such theories also
  have tradeoff relations between the predictabilities of the which-way and which-phase measurements, and these can be distinct from the quantum one. \blk
For instance, one can construct four examples of such tradeoff relations from the four foil theories presented in Ref.~\cite{ContextualityViaUR} if one takes the pair of complementary measurements described in each of these to be which-way and which-phase measurements. Under this mapping, the various forms of uncertainty relation for complementary measurements that are achieved for these different foil theories (described in Eqs.~(3)-(6) of Ref.~\cite{ContextualityViaUR}) yield corresponding forms of the wave-particle duality relation.  
\blk

\subsection{How to implement an experimental test of contextuality based on wave-particle duality}

It is also possible to implement an experimental test of contextuality via the phenomenology of wave-particle duality relations.  Here again, we leverage a result from Ref.~\cite{ContextualityViaUR}.

Any experimental test of contextuality should not presume the correctness of quantum theory, but should instead allow that some other operational theory describes the experiment.  It follows that one cannot assume that the operational theory governing the experiment satisfies the $A_1^2$-symmetry property. Instead, one simply needs to experimentally identify two measurements $M$ and $M'$ and an experimentally realizable operational state 
 that  satisfies the $A_1^2$-orbit-realizability condition relative to $M$ and $M'$ and then test the noncontextual bound on these.

The result can be summarized as follows.
\begin{proposition}\label{ContFromPredTradeoff} Consider an experimentally realizable state that satisfies the $A_1^2$-orbit-realizability condition relative to measurements $M$ and $M'$. The experiment fails to admit of a noncontextual ontological model if and only if the measured values of the $M$-predictability and the $M'$-predictability for this state violate the inequality of Eq.~\eqref{NCBoundZXUR}. 
\end{proposition}

The specialization of this result to the case where $M$ and $M'$ are the which-way and which-phase measurements, can be summarized as follows:

\begin{corollary}\label{ContFromWPduality}
Consider an experimentally realizable  state that satisfies the $A_1^2$-orbit-realizability condition relative to the which-way and which-phase measurements. The experiment fails to admit of a noncontextual ontological model if and only if the measured values of the path distinguishability $\mathcal{P}$ and the fringe visibility $\mathcal{V}$ for this state violate the inequality $\mathcal{V} + \mathcal{P}\le 1$ (Eq.~\eqref{NCBoundWPDuality}).
\end{corollary}

For concreteness, we now describe a quantum set-up that, according to the theoretical ideal, allows for the preparation of a state that  saturates  the quantum wave-particle duality relation while also satisfying the $A_1^2$-orbit-realizability condition.

Consider the preparation procedure depicted in Fig.~\ref{MeasPrep}~(c), but where the phase shift $\phi$ is limited to be 0 or $\pi$, and the reflectivity is limited to a pair of values $r$ and $1-r$ which are bounded away from 0 and 1.  
In quantum theory, these procedures are represented by the following quadruple of pure states of the dual-rail qubit $\{ \rho_i = |\psi_i\rangle \langle \psi_i|\}_{i=1,\dots,4}$, where 
\begin{align}\label{fourquantumstates}
|\psi_1\rangle &= \sqrt{r} |L\rangle + \sqrt{1-r} |R\rangle,\nonumber\\
|\psi_2\rangle &= \sqrt{r} |L\rangle - \sqrt{1-r} |R\rangle,\nonumber\\
|\psi_3\rangle &= \sqrt{1-r} |L\rangle - \sqrt{r} |R\rangle,\nonumber\\
|\psi_4\rangle &= \sqrt{1-r} |L\rangle + \sqrt{r} |R\rangle.
\end{align}
One easily verifies that these form an $A_1^2$-orbit relative to the which-way and which-phase observables.

\begin{figure}[htbp] 
\centering
\includegraphics[width=3in]{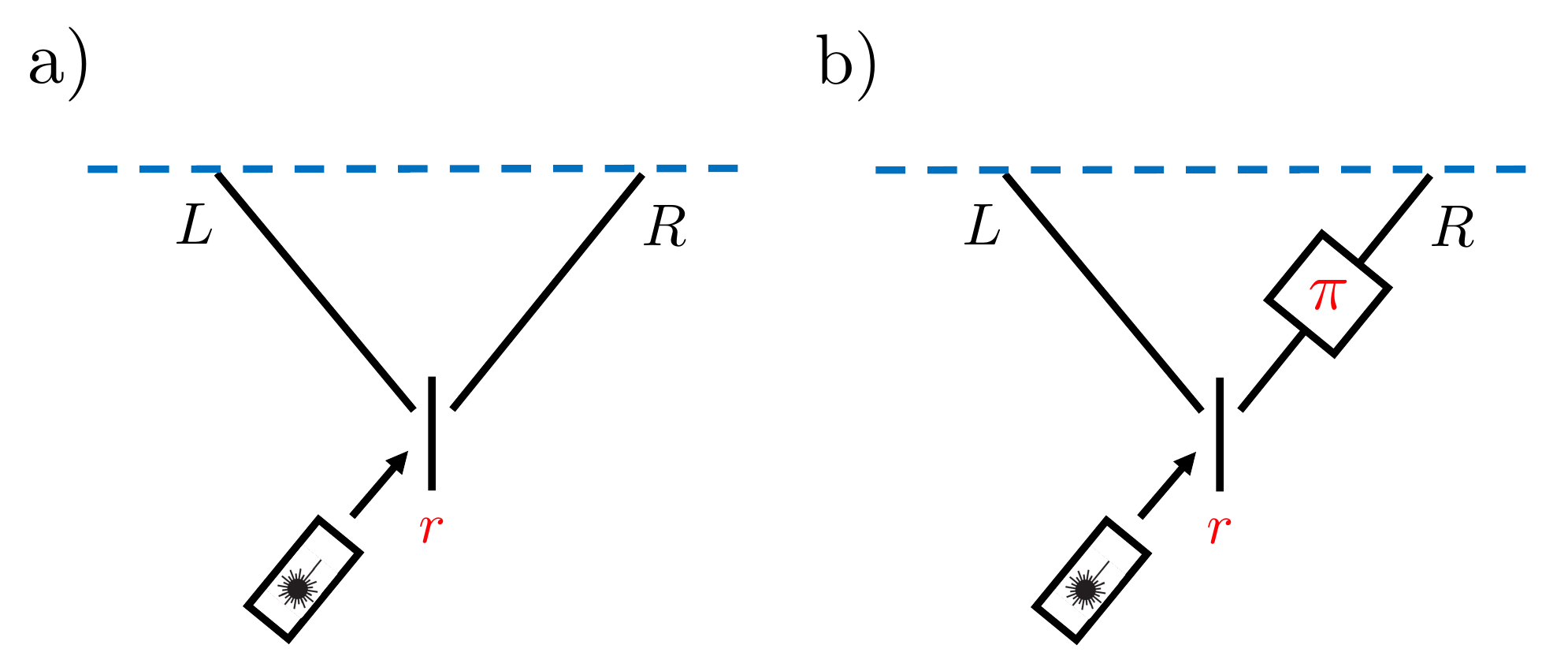}
\includegraphics[width=2.8in]{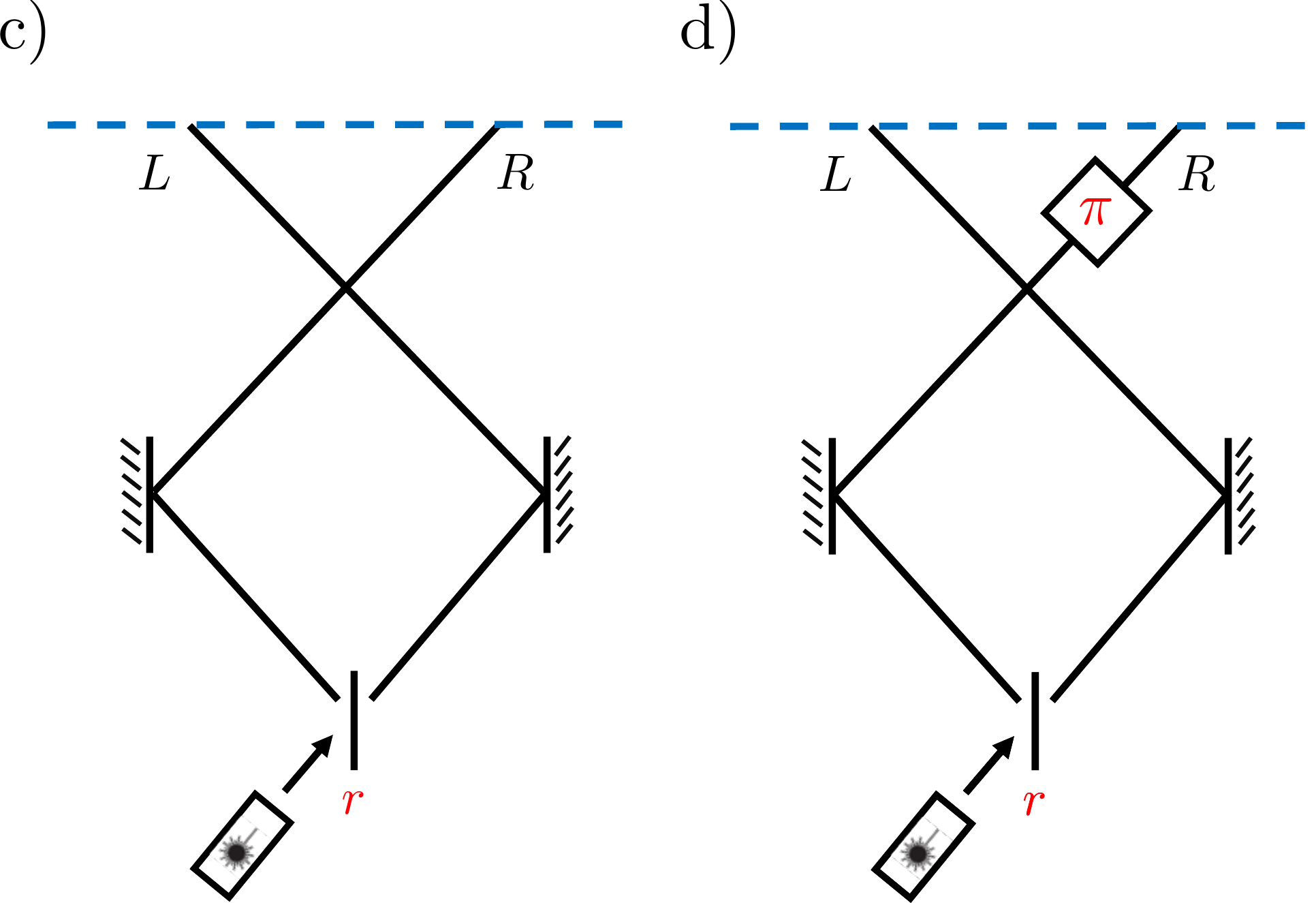}
\caption{Preparation procedures associated to each of the four pure quantum states in Eq.~\eqref{fourquantumstates}. 
}
\label{fig:OrbitOfStates}
\end{figure}

A  simple interferometric implementation of these four states is depicted in Fig.~\ref{fig:OrbitOfStates}.  Rather than requiring two distinct beamsplitters, one with reflectivity $r$ and the other with reflectivity $1-r$, one uses the reflectivity-$r$ beamsplitter to simulate the reflectivity-$(1-r)$ beamsplitter by following it up with a swap operation between the modes.

In such an experiment, one is aiming to implement the theoretical ideal of each of  the four preparation procedures described in Fig.~\ref{fig:OrbitOfStates} as well as the theoretical ideal of the which-way measurement and the which-phase measurement depicted in Figs.~\ref{MeasPrep}~(a) and b) respectively.  However, the idealization of purity and sharpness 
  is never realized in practice.  Rather, one realizes noisy and inaccurate versions of the pure states of Eq.~\eqref{fourquantumstates} and of the sharp measurements $\{ |L\rangle, |R\rangle\}$ and $\{ |+\rangle, |-\rangle\}$.  It follows that one must characterize precisely which states and effects are realized in the experiment.  This can be achieved using the tomography scheme of Refs.~\cite{Mazurek2021,bootstrap2}.  

The latter scheme does not presume the correctness of quantum theory, but rather fits the data to states and effects in the framework of generalized probabilistic theories.  This aspect of the scheme---that it is {\em theory-agnostic}---enables one to adjudicate between quantum theory and alternative operational theories, such as those that {\em do} admit of a noncontextual model \rob or those that manifest greater inequality violations than quantum theory permits. \blk In this way, one obtains real-valued vector representations of the two realized measurements, as well as real-valued vector representations $\vec{s}_1, \vec{s}_2, \vec{s}_3, \vec{s}_4$ of the realized preparation procedures.\footnote{Without presuming the correctness of quantum theory, one cannot know in advance what sets of preparations and measurements are tomographically complete, so the plausibility of the hypothesis of completeness depends on how much work has been done in attempting to falsify it. The fact that the hypothesis could in principle be falsified in the future by some exotic laboratory procedure is the most significant loophole in experimental tests of noncontextuality. See the introduction of Ref.~\cite{Mazurek2021,bootstrap2} or Refs.~\cite{PuseydelRio,NCsubsystems} for more discussion of this point.}

\rob The four states that are actually realized in the experiment, by virtue of being noisy and inaccurate versions of the theoretical ideals, will generally only {\em approximately} satisfy the  $A_1^2$-orbit-realizability condition relative to the realized measurements (Eqs.~\eqref{eq:rectanglesymmetryEV} and \eqref{eq:op_equiv}).  
However, one can always define a quadruple of {\em secondary states} lying in the convex hull of those that {\em were} realized, such that these {\em do} satisfy the condition.  Every state that is a convex mixture of the realized states must be included in the operational theory, since the state space is presumed to be convexly closed.  As a result, these secondary states are known to be included in the set that are experimentally realizable.  \blk  Hence, even though the states that were actually realized in the experiment may not satisfy the $A_1^2$-orbit-realizability condition exactly, one can always define secondary states that {\em do}, and if {\em these} violate the noncontextual bound, one has witnessed the fact that the experimental data cannot be explained by a noncontextual model.  (See Ref.~\cite{Mazurek2016} for a discussion of the technique of secondary procedures.) \blk
 
Evidently, choosing the reflectivity $r=3/4$ for the beamsplitter in the preparation stage provides the best opportunity for witnessing contextuality, as this is the value that, in the ideal version of the experiment, leads to the largest violation of the noncontextual bound.

\section{Discussion}

As seen in Fig.~\ref{QTradeoffRelationVsNCBound}, the quantum tradeoff curve intersects the noncontextual bound at $(\mathcal{V}, \mathcal{P})=(1,0)$ and $(\mathcal{V}, \mathcal{P})=(0,1)$.  As noted earlier, the $(\mathcal{V}, \mathcal{P})=(1,0)$ point corresponds to the preparation of a which-phase eigenstate and the $(\mathcal{V}, \mathcal{P})=(0,1)$ point corresponds to the preparation of a which-way eigenstate. These are the only cases that are typically referenced by researchers who claim that interference phenomena capture the essence of quantum theory.   In other words, they are the only cases that appear in the TRAP phenomenology. In Ref.~\cite{ToyFieldTheory}, it was shown that these cases admit of a  noncontextual ontological model, thereby undermining the conventional claim.   
The present article provides an alternative way to see that this is the case,  and hence provides an alternative way of undermining the conventional claim. 

The TRAP phenomenology of interference is fully captured by  preparations of the dual-rail qubit that only make use of a 50-50 beamsplitter and relative phase shifts that only take values  0 or $\pi$ (together with the which-way and which-phase measurements). 
For any departure from this class of preparation procedures, the question arises of whether one can provide a classical explanation of it.   
Of course, to vindicate the claim that a particular experimental set-up resists classical explanation, one must prove a no-go theorem, a methodological point that was emphasized in Sec.~V.B.4 of Ref.~\cite{ToyFieldTheory}.

We noted in the introduction that it is actually {\em straightforward} to prove a noncontextuality no-go theorem using a dual-rail qubit.  It suffices to translate into the concrete framework of a dual-rail qubit {\em any} of the existing proofs based a single qubit. \rob There are many to choose from.\footnote{\label{qubitproofs}Those that leverage only operational equivalences among the preparations include the first proof of the failure of generalized noncontextuality, based on states forming a star of David in the Bloch sphere~\cite{Spekkens2005}, the proofs based on the probability of success in 2-bit or 3-bit parity-oblivious multiplexing~\cite{POM}, the proof based on the probability of error-free state discrimination of a pair of nonorthogonal states~\cite{SchmidSpekkens2018}, the proof based on the probability of unambiguous discrimination of a pair of nonorthogonal states~\cite{Flatt2021}, and the proof based on the probability of success in probabilistic cloning of a pair of nonorthogonal states~\cite{lostaglio2020contextual}.  \blk In addition, there are those that also leverage operational equivalences among the measurements. These include the first proof based an assumption of noncontextuality for unsharp measurements, given in Sec. V of Ref.~\cite{Spekkens2005}, which was also used in Ref.~\cite{Mazurek2016},
and the proof based on three unsharp measurements that are jointly measurable pairwise but not triplewise~\cite{Kunjwal2014}. 
There are even proofs that leverage operational equivalences among transformations.  These include the original proof of the failure of transformation noncontextuality~\cite{Spekkens2005}, the proof arising from different decompositions of the completely depolarizing channel in terms of Clifford unitaries~\cite{lillystone2019contextuality}, 
and proofs based on the transformative aspect of measurements, which concern pre- and post-selected scenarios~\cite{Pusey2014,pusey2015logical,KunjwalLostaglioPusey2019}. 
 \blk}
Furthermore, because many of these only make use of states and effects in a {\em single plane} of the Bloch sphere\footnote{Of the ones listed in footnote \ref{qubitproofs}, the only ones that are {\em not} of this type, i.e., the only ones that extend out of a single plane of the Bloch sphere, are the 3-bit scheme of Ref.~\cite{POM}, and the proofs in Refs.~\cite{Kunjwal2014} and \cite{lillystone2019contextuality}. \blk}
  \blk
  it follows that simple modifications of the experimental set-up appearing in the TRAP phenomenology are sufficient to obtain a no-go result. 

For instance, if, in the preparation stage, one keeps the reflectivity of the beamsplitter at $\frac12$ (as is the case for the TRAP phenomenology) but one allows relative phase shifts in the full region $[0,2\pi]$ rather than simply 0 or $\pi$, then this is sufficient to prepare any state in the $\hat{x}{-}\hat{y}$ plane of the Bloch sphere (relative to the association of Pauli-$Z$ and Pauli-$X$ with the bases $\{|L\rangle, |R\rangle\}$ and $\{(1/\sqrt{2})(|L\rangle +|R\rangle),(1/\sqrt{2})(|L\rangle - |R\rangle)\}$).  If one has more than a single measurement in this plane, for instance, one allows measurements of the bases  $\{(1/\sqrt{2})(|L\rangle +|R\rangle),(1/\sqrt{2})(|L\rangle - |R\rangle)\}$ and $\{(1/\sqrt{2})(|L\rangle + i|R\rangle),(1/\sqrt{2})(|L\rangle - i|R\rangle)\}$, \rob that is, both the Pauli-$X$ and Pauli-$Y$ observables, \blk
  then one can derive a no-go result.  
However, such a result cannot obviously be interpreted in terms of wave-particle duality.  Indeed, it is not clear whether it would have any significance in terms of aspects of the phenomenology of interference that were of prior interest to researchers. 

But now consider the case where, in the preparation stage, the relative phase shift is still constrained to be 0 or $\pi$ (as is the case for the TRAP phenomenology), but the reflectivity of the beamsplitter is allowed to be anything in the range $[0,1]$.  
This is also sufficient to prepare any state in the plane of the Bloch sphere spanned by the eigenstates of the which-way and which-phase observables (i.e., the $\hat{x}{-}\hat{z}$ plane).  Together with the which-way and which-phase measurements, this is sufficient to derive a no-go result.   The no-go result described in this article is an instance of this case.  Because it refers to 
the quantum tradeoff between the fringe visibility and the path distinguishability, which {\em has} been previously thought to be significant in the context of interferometry, it satisfies the desideratum we articulated in the introduction.

\blk
Recent work by Wagner {\em et al.}~\cite{Wagner2022a} also considered what aspects of the phenomenology of interference might resist explanation in terms of a noncontextual ontological model.  
In particular, they translate the qubit proofs of contextuality 
of Refs.~\cite{Wagner2020,Wagner2022b} into the language of a dual-rail qubit.  This no-go result does not (as far as we can tell) directly relate to phenomena that have been previously thought to be significant in the context of interferometry. Nonetheless, Wagner {\em et al.} also demonstrate a no-go result based on the probability of succeeding in the task of {\em quantum interrogation}, which {\em was} of prior interest in the interferometric setting. Quantum interrogation is a generalization of the bomb-testing task considered by Elitzur and Vaidman~\cite{ElitzurVaidman1993} wherein one seeks to maximize the probability of detecting the presence of a detector without causing it to fire (or of detecting a bomb without exploding it).
The aspect of the task that Wagner {\em et al.} show can witness contextuality is the dependence of the efficiency (a function of the probability of success in the task) on a parameter describing the confusability of the states (related to the nonorthogonality of the states in the quantum case). 
Since the probability of success in this task has been of prior interest in the context of interferometry, this result provides the type of answer to the question of ``what is nonclassical about the phenomenology of interference?''  that we are looking for. It is a complementary answer to the one we have given here.

Given that there has already been an experiment showing that the quantum wave-particle duality relation considered here (wherein the which-way and which-phase measurements are counterfactual alternatives) can be saturated to good approximation~\cite{liu2012relation},
 the question naturally arises whether this experiment
  already demonstrated  the nonclassical aspects of the phenomenology of interference highlighted in this article. The answer is: not quite.  Although it did confirm
    that there exist states that can approximately saturate the ideal quantum tradeoff, it
     did not verify that these states satisfy the $A_1^2$-orbit realizability condition.  Only if the latter condition is satisfied is one justified in inferring from the assumption of noncontextuality that the tradeoff is bounded by a linear curve. 
Furthermore, in order to allow for the possibility of a failure of operational quantum theory in the experiment \rob (for instance, to allow for the possibility of violations of the noncontextual bound {\em greater} than what quantum theory allows), \blk it is important to characterize the preparations and measurements using the technique of theory-agnostic tomography~\cite{Mazurek2021,bootstrap2}.
It is relative to these characterizations that one can identify a quadruple of preparations that satisfy the  $A_1^2$-orbit-realizability condition,
 and then test whether these yield values of fringe visibility $\mathcal{V}$ and path distinguishability $\mathcal{P}$ that violate the noncontextuality inequality $\mathcal{V}+\mathcal{P}\le 1$.
In this way, one can implement a direct experimental test of contextuality via the wave-particle duality relation. This relation, therefore, is an aspect of the phenomenology of interference that can witness nonclassicality.

\section*{Acknowledgments}
This research was supported by Perimeter Institute for Theoretical Physics. Research at Perimeter Institute is supported by the Government of Canada through the Department of Innovation, Science and Economic Development Canada and by the Province of Ontario through the Ministry of Research, Innovation and Science.  RWS was also supported by the Natural Sciences and Engineering Research Council of Canada (Grant No. RGPIN-2017-04383). 
ML was supported in part by the Fetzer
Franklin Fund of the John E. Fetzer Memorial Trust and
by grant number FQXi-RFP-IPW-1905 from the Foundational Questions Institute and Fetzer Franklin Fund, a donor advised fund of Silicon Alley Community Foundation.
GSc is supported by QuantERA/2/2020, an ERA-Net co-fund in Quantum Technologies, under the eDICT project. LC acknowledges funding from the Einstein Research Unit `Perspectives of a Quantum Digital Transformation'.
DS acknowledge support by the Foundation for Polish Science (IRAP project, ICTQT, contract no.2018/MAB/5, co-financed by EU within Smart Growth Operational Programme). GS is thankful to the group QUANTUM from Bari for the invaluable scientific and human support they provided during the early stage of this project.

\bibliography{bib_new}

\end{document}